\documentclass[aps,preprint,nofootinbib,showpacs,superscriptaddress]{revtex4}
\usepackage{amssymb}
\usepackage{amsmath}
\usepackage{graphicx}
\usepackage{multirow}
\usepackage{subfigure}
\usepackage{float}
\usepackage{graphics}
\usepackage{slashed}

\def\ra{\rightarrow}
\def\be{\begin{equation}}
\def\ee{\end{equation}}
\newcommand{\bea}{\begin{equation} \begin{array}{c}}
\newcommand{\eea}{ \end{array} \end{equation}}

\def\as{\alpha_s}

\def\nno{\nonumber}

\newcommand{\M}{\mathcal{M}}

\begin{document}

\title{Next-to-leading order QCD effect of $W'$
  on top quark Forward-Backward Asymmetry }
\author{Kai Yan}
\affiliation{Department of Physics and State Key Laboratory of
Nuclear Physics and Technology, Peking University, Beijing 100871,
China}
\author{Jian Wang}
\affiliation{Department of Physics and State Key Laboratory of
Nuclear Physics and Technology, Peking University, Beijing 100871,
China}
\author{Ding Yu Shao}
\affiliation{Department of Physics and State Key Laboratory of
Nuclear Physics and Technology, Peking University, Beijing 100871,
China}
\author{Chong Sheng Li}
\email{csli@pku.edu.cn} \affiliation{Department of Physics and State
Key Laboratory of Nuclear Physics and Technology, Peking University,
Beijing 100871, China} \affiliation{Center for High Energy Physics,
Peking University, Beijing 100871, China}

\begin{abstract}
We present the calculations of the complete next-to-leading order (NLO) QCD
corrections to the total cross section, invariant mass distribution and
the forward-backward asymmetry ($\rm A_{FB}$) of top quark pair production mediated by $W'$ boson.
Our results show that in the best fit point in the parameter space allowed by data at the Tevatron,
the NLO corrections change the new physics contributions to the total cross section slightly,
but increase the $\rm A_{FB}$ in the large invariant mass region by about $9\%$.
Moreover, we evaluate the total cross section  and charge asymmetry ($\rm{A}_{\rm{C}}$) of top pair production
at the LHC, and find that both total cross section and $A_{\rm C}$ can be used to distinguish NP from SM with the integrated luminosity increasing.
\end{abstract}


\maketitle

\section{INTRODUCTION}\label{s1}
The top quark is the heaviest particle discovered so far,
with a mass close to the electroweak symmetry breaking scale.
Thus it is a wonderful probe for the electroweak breaking mechanism and
new physics (NP) beyond the standard model (SM) through its productions and decays at colliders.
The forward-backward asymmetry ($\rm A_{FB}$) of the top quark pair production is
one of the interesting observables in the top quark sector. Within the SM,
$\rm A_{FB}$ is absent at the tree level in QCD due to charge
symmetry, and occur at QCD next-to-leading order (NLO) with the
prediction $\rm A_{FB} \sim 6\%$ in the $t\bar{t}$ rest frame
\cite{Kuhn:1998jr,Kuhn:1998kw,Bowen:2005ap,Almeida:2008ug,Antunano:2007da,Ahrens:2011uf}.
In the last few years, D\O ~and CDF Collaborations have measured
$\rm A_{FB}$ at the Tevatron
\cite{:2007qb,Aaltonen:2008hc,Aaltonen:2011kc,collaboration:2011gf}.
Recently, the CDF Collaborations annouced that, for the invariant
mass of the top quark pair $m_{t\bar t}\geq 450$~GeV, the measured
asymmetry in the $t\bar{t}$ rest frame is $\rm
A_{FB}=0.475\pm0.114$\cite{Aaltonen:2011kc}, which differs by
3.4$\sigma$ from the SM predictions $\rm A_{FB}=0.088\pm0.013$. This
deviation has stimulated a number of theoretical papers on NP
models, such as new gauge bosons,
axigluons\cite{Djouadi:2009nb,Jung:2009jz,Cheung:2009ch,Frampton:2009rk,
Shu:2009xf,Arhrib:2009hu,Ferrario:2009ee,Dorsner:2009mq,Jung:2009pi,
Cao:2009uz,Barger:2010mw,Cao:2010zb,Xiao:2010hm,Martynov:2010ed,
Chivukula:2010fk,Bauer:2010iq,Chen:2010hm,Jung:2010yn,Burdman:2010gr,
Jung:2010ri,Choudhury:2010cd,Cheung:2011qa,Cao:2011ew,Berger:2011ua,
Barger:2011ih,Bhattacherjee:2011nr,Blum:2011up,Patel:2011eh,
Isidori:2011dp,Zerwekh:2011wf,Barreto:2011au,Foot:2011xu,Ligeti:2011vt,
Gresham:2011pa,Jung:2011zv,Buckley:2011vc,Shu:2011au,AguilarSaavedra:2011zy,
Chen:2011mga,Degrande:2011rt,Jung:2011ua,Jung:2011ue,
Babu:2011yw,Djouadi:2011aj,Barcelo:2011fw,Krohn:2011tw,AguilarSaavedra:2011hz,
Cui:2011xy,Hektor:2011ms,Gabrielli:2011jf,Duraisamy:2011pt,
AguilarSaavedra:2011ug,Barcelo:2011vk,Tavares:2011zg,Vecchi:2011ab,Blum:2011fa,
Degrande:2010kt,AguilarSaavedra:2011vw,Shao:2011wa,Ko:2011vd,Frank:2011rb,
Davoudiasl:2011tv,Jung:2011id,Westhoff:2011tq,Krnjaic:2011ub,Berger:2011sv,
AguilarSaavedra:2011cp,Cao:2011hr,Kuhn:2011ri,Endo:2011tp,Kolodziej:2011ir}.

Recent studies are concerned with the problem of top asymmetry  by a
flavor-changing interaction mediated by a charged vector boson, $W'$~\cite{Cheung:2009ch,PhysRevD.81.014016},
which can be described by the following effective Lagrangian~\cite{Cheung:2009ch}:
\begin{eqnarray}
  \mathcal {L}_{NP}= -g'W'^{-}_{\mu}\bar{d}\gamma^{\mu}(f_{L}P_{L}+f_{R}P_{R})t+h.c.,
\end{eqnarray}
where $P_{R,L}=(1\pm\gamma^{5})/2$ are the chirality projection operators,
$f_{L,R}$ are the chiral couplings of the $W'$ boson with fermions, satisfying $f_{L}^2+f_{R}^2=1$,
and $g'$ is the coupling constant.
The study of this model at the leading order (LO) has been explored in Refs.~\cite{Cheung:2011qa,PhysRevD.81.114004}.
It is shown that, for suitable parameters, this model can explain the $\rm A_{FB}$ observed at the Tevatron within 1-1.5$\sigma$ of the data.
It is well known that the LO cross sections for process at hardron colliders suffer form large uncertainties due to
the arbitrary choice of the renormalization and factorization scales, thus it is necessary to include higher order corrections to make a reliable theoretical prediction.
Besides, at the NLO level, virtual corrections, real gluon emission and massless (anti)quark emission can lead to a sizeable difference between the differential top and
anti-top production process~\cite{Kuhn:1998jr,Kuhn:1998kw}, which will also contribute to $\rm A_{FB}$.
Therefore it is necessary to  perform
complete calculations of  NLO contributions in the $W'$ model.

There is a similar work in the $Z'$ model~\cite{PhysRevD.82.034026},
where the NLO QCD corrections up to $\mathcal{O}(\alpha_s^2 {g'}^2)$ are taken into account.
In this work, we calculate both $\mathcal{O}(\alpha_s^2 g'^2) $ and $ \mathcal{O}(\alpha_s g'^4)$ NP contributions,
and the latter term is not definitely smaller than the former so that it should not be neglected.
Based on the above calculation, we fit the data at the Tevatron, including total cross section, the invariant mass distribution and the $\rm A_{FB}$, and find the allowed parameter space.
Moreover, we study the top quark pair production at the Large Hadron Collider (LHC) induced by a $W'$ boson at the NLO QCD level.
Since the gluon fusion channel dominates in the $t \bar {t} $ production process at the LHC,
it is difficult to probe NP effects on $\rm A_{\rm FB}$  from early LHC results.
However, LHC will be able to detect the potential NP effect on the Charge Asymmetry($\rm A_{\rm C}$)
when the integrated luminosity increases in future.

The arrangement of this paper is as follows.
In section \ref{s2} we show the LO results of top quark pair production.
In section \ref{s3}, we present the details of the NLO calculations,
including the virtual and real corrections.
In section \ref{s4} we show the numerical results. Conclusion is given in section~\ref{s5}.

\section{Leading order results}\label{s2}
Up to NLO, the $t \bar t$ production amplitudes, including NP contributions, can
be written as
\begin{eqnarray}
\M ^{t \bar t} =\alpha_{s} f_{\rm SM}^{\rm LO}+ g'^2 f_{\rm NP}^{\rm LO} +\alpha_{s}^2 f_{\rm SM}^{\rm NLO}+ \alpha_{s}g'^2 f_{\rm NP}^{\rm NLO}.
\end{eqnarray}
And the $t \bar t$ amplitude squared is
\begin{eqnarray}
|\M^{t\bar t}|^2&=&\alpha_{s}^2  f_{\rm SM}^{\rm LO} f_{\rm SM}^{\rm LO^\ast}+2\alpha_{s} g'^2 \mathcal{R}e \left( f_{\rm SM}^{\rm LO} f_{\rm NP}^{\rm LO^\ast}\right)+g'^4  f_{\rm NP}^{\rm LO} f_{\rm NP}^{\rm LO^\ast}\nno\\
&+& 2 \alpha_{s}^3 \mathcal{R}e \left(f_{\rm SM}^{\rm LO} f_{\rm SM}^{\rm NLO^\ast} \right)+2 \alpha_{s}^2g'^2\left[ \mathcal{R}e \left(f_{\rm SM}^{\rm LO} f_{\rm NP}^{\rm NLO^\ast} \right)+ \mathcal{R}e \left(f_{\rm NP}^{\rm LO} f_{\rm SM}^{\rm NLO^\ast} \right)\right]\nno\\
&+&2 \alpha_{s}g'^4 \mathcal{R}e \left(f_{\rm NP}^{\rm LO} f_{\rm NP}^{\rm NLO^\ast} \right).
\end{eqnarray}
In the first line is the LO result of SM and NP.
The first term in the second line is the SM NLO result and the second term is the interference contribution.
The NP NLO result is given in the third line.
\begin{figure}
  \includegraphics[width=0.6\linewidth]{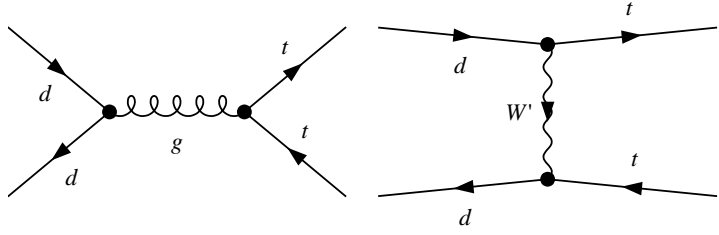}\\
  \caption{LO Feynman diagrams for $d\bar{d}\rightarrow t\bar{t}$.}
\label{fig:FeynDiaLO}
\end{figure}

The LO Feynman diagrams for the subprocess $d(p_1)\bar d(p_2)\rightarrow t(p_3)\bar t(p_4)$ induced by the NP and the SM
QCD interactions are shown in Fig.~\ref{fig:FeynDiaLO},
and the LO  partonic cross section can be written as
\begin{equation}
\hat \sigma^{\rm LO}=\hat \sigma^{\rm LO}_{\rm SM}+\hat \sigma^{\rm LO}_{\rm {INT}}+\hat \sigma^{\rm LO}_{\rm NPS}
\end{equation}
where subscripts SM, INT and NPS denote the SM channel contributions,
the interference between SM and NP channels, and NP channel contributions, respectively.
The LO partonic differential cross sections are given by
\begin{eqnarray}
\frac{ d\hat{\sigma}^{\rm LO}_{\rm SM} }{d\cos \theta }&=& \frac{2 \pi \beta}{9\hat{s}}\frac{\alpha _s^2}{\hat{s}^2}\big[6 m_{t}^{4}-4 m_{t}^{2} (t+u)+t^{2}+u^{2} \big],
\end{eqnarray}
\begin{eqnarray}
\frac{ d\hat{\sigma}^{\rm LO} _{\rm INT} }{d\cos \theta }&=&\frac{2 \beta }{9 \hat {s}} \frac{\alpha_{s} g'^2}{m_{\rm W'}^2 s \left(m_{\rm W'}^2-t\right)} \left(f_{R}^2+f_{L}^2\right)\big[3 m_{t}^6+m_{t}^4 \left(6 m_{\rm W'}^2-3 t-u\right)\nno\\
&& +m_{t}^2 \left(t^2-2 m_{W'}^2 (t+3 u)\right)+2 m_{W'}^2 u^2\big],
 \end{eqnarray}
 \begin{eqnarray}
 \frac{ d\hat{\sigma}^{\rm LO}_{\rm NPS} }{d\cos \theta }&=& \frac{\beta}{8 \pi  \hat{s}} \frac{g'^4}{ m_{\rm W'}^{4} (m_{\rm W'}^2-t)^{2}}
  \big\lbrace (f_{R}^4+f_{L}^4) \big[m_{t}^8-2  m_{ t}^6 t \nno\\
  &&+ m_{ t}^4 \left(4 m_{\rm W'}^4+4  m_{\rm W'}^2 s+t^2\right)-8 m_{t}^2  m_{\rm W'}^4 u+4 m_{\rm W'}^4 u^2\big]\nno\\
  &&+2 f_{R}^2 f_{L}^2 \left[m_{t}^8-2 m_{t}^6 t+m_{t}^4 \left(4  m_{\rm W'}^2 s+t^2\right)-8 m_{\rm W'}^2 m_{\rm W'}^4 s+4  m_{\rm W'}^4 s^2\right]\big\rbrace,
 \end{eqnarray}
where the Mandelstam variables $s$, $t$ and $u$ are defined as follows:
\begin{eqnarray}
  s=(p_1 + p_2)^2,~~~t=(p_1 - p_3)^2,~~~u = (p_1 - p_4)^2.
\end{eqnarray}
The relations between them are
\begin{eqnarray}
m_{t}^2-t=s(1-\beta \cos \theta)/2, ~~~~~m_{t}^2-u=s(1+\beta \cos \theta)/2,
\end{eqnarray}
where $\beta=\sqrt{1-4m_t^2/s}$, and $\theta$ is the
polar angle of the outgoing top quark in the $t\bar t$ rest frame.
The colors and spins of the incoming (outgoing) particles have been averaged (summed) over.
Integrating over $\cos \theta$, we obtain the LO result of $d\bar{d}\to t\bar{t}$ partonic cross section.

The LO total cross section at the hadron collider is
obtained by convoluting the partonic cross section with the parton distribution functions (PDF) $G_{i/A(B)}$ for the initial hadrons A and B:
\begin{eqnarray}
  \sigma^{\rm{LO}}=\sum_{a,b}\int_{\tau}^1dx_a\int_{\tau/x_a}^1dx_b
  G_{a/A}(x_a,\mu_f)G_{b/B}(x_b,\mu_f)\hat{\sigma}^{\rm{LO}},
\end{eqnarray}
where $\tau=4m_t^2/s$.

\section{Next-to-leading order QCD corrections}\label{s3}
The NLO corrections to the top quark pair production consist of the
virtual corrections, generated by loop diagrams of colored
particles, and the real corrections with the radiation of a real
gluon or a massless (anti)quark.
For the real corrections, we used the two cutoff phase space slicing method
 to subtract the infrared (IR) divergences\cite{PhysRevD.65.094032}.

\subsection{Virtual corrections}
\begin{figure}
\includegraphics[scale=1]{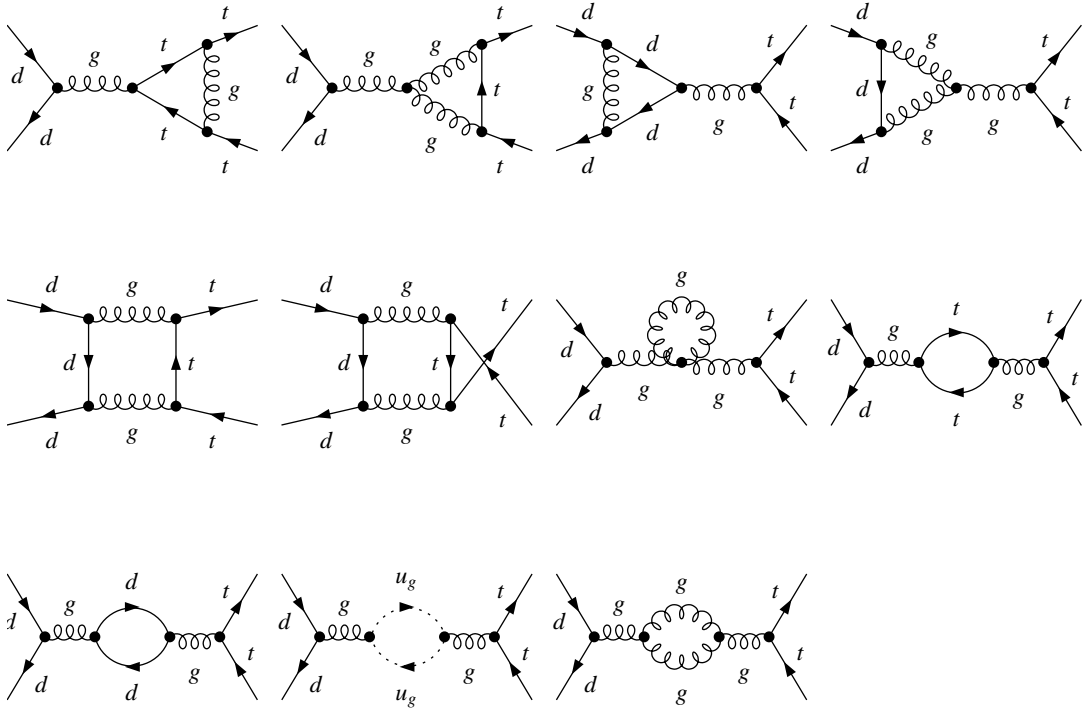}
\caption{One-loop virtual Feynman diagrams for
$d\bar{d}\rightarrow t\bar{t}$ induced by SM QCD interactions. }
\label{fig:smloop}
\end{figure}

\begin{figure}
\includegraphics[scale=1]{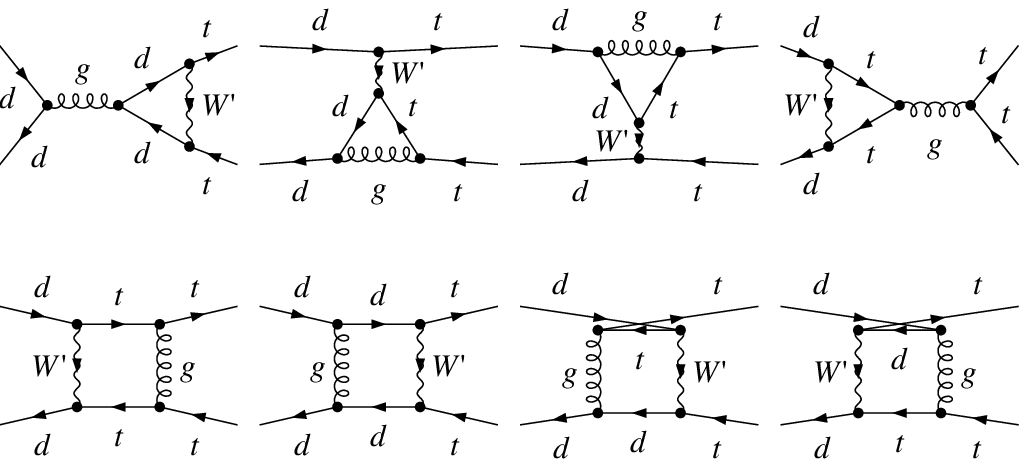}
\caption{One-loop virtual Feynman diagrams for
$d\bar{d}\rightarrow t\bar{t}$ induced by NP interactions. }
\label{fig:NPloop}
\end{figure}

The virtual corrections for the top quark pair production include the box diagrams, triangle diagrams,
and self-energy diagrams induced by SM QCD and NP interactions as shown in Fig.~\ref{fig:smloop} and Fig.~\ref{fig:NPloop}, respectively.
The renormalized virtual amplitudes are given as follows
\begin{eqnarray}
  \M^{ren}_{\rm SM} &=& \frac{\alpha_{s}}{4 \pi} C_{\epsilon} C_{F}  \left( \frac{2}{\epsilon_{\rm UV}} \right)\M^{\rm LO}_{\rm SM}+\left( \delta Z_{2}^{q}+ \delta Z_{2}^{t}+ 2 \delta g_{s} \right)\M^{\rm LO}_{\rm SM}+\M^{fin}_{\rm SM},\\
  \M^{ren}_{\rm NP} &=& \frac{ \alpha_{s}}{4 \pi} C_{\epsilon} C_{F} \left( \frac{2}{\epsilon_{\rm UV}} \right)\M^{\rm LO}_{\rm NP}+\left( \delta Z_{2}^{q}+ \delta Z_{2}^{t} \right)\M^{\rm LO}_{\rm NP}+  \M^{fin}_{\rm NP},
\end{eqnarray}
where $C_{\epsilon}=(4\pi)^{\epsilon}\frac{1}{\Gamma(1-\varepsilon)}$.
$ \M^{fin}_{\rm SM} $ and $\M^{fin}_{\rm NP} $ are ultraviolet(UV) finite terms for SM and NP processes.
All the UV divergences in the loop diagrams are canceled by counterterms
$\delta Z_{2}^{q}$ for the wave functions of the external fields in on-shell scheme,
and $\delta g_{s}$ for the strong coupling constant in the $\rm \overline{MS}$ scheme modified to decouple the top quark\cite{PhysRevD.18.242},
\begin{eqnarray}
 \delta Z_{2}^{q} &=& -\frac{\alpha_{s}}{3\pi} C_{\epsilon} \left(\frac{1}{\epsilon_{\rm UV}}-\frac{1}{\epsilon_{\rm IR}}\right),\\
 \delta Z_{2}^{t} &=& -\frac{\alpha_{s}}{3\pi} C_{\epsilon} \left( \frac{1}{\epsilon_{\rm UV}}+\frac{2}{\epsilon_{\rm IR}}+4+3\ln \frac{\mu_r ^{2}}{m_t^{2}} \right),\\
 \delta g_{s} &=& \frac{\alpha_{s}}{4\pi} C_{\epsilon} \left( \frac{n_{f}}{3}-\frac{11}{2} \right) +\frac{\alpha_{s}}{12\pi}C_{\epsilon} \left(\frac{1}{\epsilon_{\rm UV}}+ \ln \frac{\mu_r^{2}}{m_{t}^{2}} \right),
\end{eqnarray}
where $n_f=5$ and $\mu_{r}$ is the renormalization scale.
The renormalized amplitudes $\M^{ren}_{\rm SM}$ and $\M^{ren}_{\rm NP}$  are UV finite, but still contains IR divergences.
The  virtual corrections for subprocess $q\bar{q} \to t\bar{t}$ can be expressed as:
\begin{eqnarray}
d\hat{\sigma}^{virt} &=& d\hat{\sigma}_{\rm SM}^{virt}+ d\hat{\sigma}_{\rm NPS}^{virt}+ d\hat{\sigma}^{virt}_{\rm INT} \nno\\
   &=& \frac{1}{2s} d\Gamma_{2}  \big\lbrace  2 \mathcal{R}e\left(\M_{\rm SM}^{ren} \M_{\rm SM}^{\rm LO\ast}\right)+ 2 \mathcal{R}e\left(\M_{\rm NP}^{ren}\M_{\rm NP}^{\rm LO\ast}\right)\nno\\
   &+& 2 \mathcal{R}e\left(\M_{\rm SM}^{ren} \M_{\rm NP}^{\rm LO \ast}+ \M_{\rm NP}^{ren} \M_{\rm SM}^{\rm LO\ast}\right) \big\rbrace.
\end{eqnarray}
We have calculated the SM contribution, and find the result agrees with that in the Ref.~\cite{Zhu:2011gd}.
The one-loop correction for the cross section induced by NP interactions, with  IR singularities separated from finite terms, is given by
\begin{eqnarray}\label{eqv}
 d\hat{\sigma}^{virt}&=& \frac{\alpha_{s}}{2 \pi} C_{\epsilon} \left[ \frac{(A_{2}^{v})_{\rm INT}}{\epsilon_{\rm IR}^{2}}+\frac{ (A_{1}^{v})_{\rm INT}}{\epsilon_{\rm IR}}\right] d \hat{\sigma}^{LO}_{\rm INT}\\&+& \frac{\alpha_{s}}{2 \pi} C_{\epsilon} \left[ \frac{(A_{2}^{v})_{\rm NPS}}{\epsilon_{\rm IR}^{2}}+\frac{ ( A_{1}^{v})_{\rm NPS}}{\epsilon_{\rm IR}}\right] d \hat{\sigma}^{\rm LO}_{\rm NPS}+d \hat{\sigma}^{virt,fin}
\end{eqnarray}
where
\begin{eqnarray}\label{eqavint}
(A_{2}^{v})_{\rm INT}&=& -2 C_{F},\\
 (A_{1}^{v})_{\rm INT} &=& \frac{C_{F}} {4} \Bigg[  16 \ln \frac{-t_{1}} {\mu_{r} ^{2}} +2 \ln \frac{-u_{1}} {\mu_{r} ^{2}}
 +9 \ln \frac{\mu_{r} ^{2}} {m_{t}^{2}}  + \ln \frac{\mu_{r} ^{2}} {s}\nno\\
 &-& \frac{ 1+\beta ^{2}} {2 \beta} \ln \frac{\beta+1} {1-\beta} -20 \Bigg],
\end{eqnarray}
and
\begin{eqnarray}\label{eqavnps}
 ( A_{2}^{v})_{\rm NPS} &=& -2 C_{F},\\
 (A_{1}^{v})_{\rm NPS} &=& 2C_{F}\left[ 2\ln \frac{-t_1}{\mu_{r} ^2} + \ln \frac{\mu_{r} ^2}{m_{t}^{2}} -\frac{5}{2}\right],
\end{eqnarray}
with $t_1 = t - m_t^2 $, $u_1 = u - m_t^2 $.
The IR divergent terms are proportional to  the LO partonic crosss section $\hat{\sigma}^{\rm LO}_{\rm INT}$
and $\hat{\sigma}_{\rm NPS}^{\rm LO}$.
$\sigma^{virt,fin}$ is the finite terms of the virtual cross section.

\subsection{Real corrections}\label{ss1}
\begin{figure}[h!]
\includegraphics[scale=0.7]{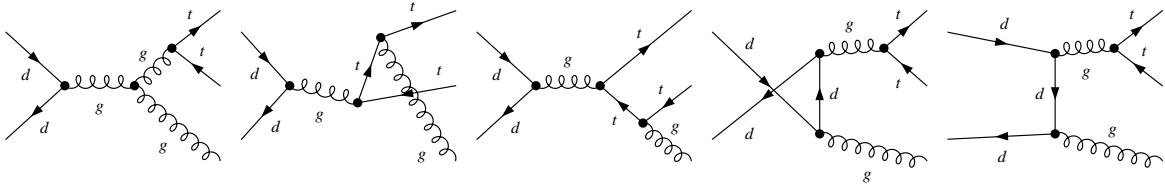}
\caption{Feynman diagrams for a gluon emission induced by SM QCD interactions.
The diagrams for a (anti)quark emission can be obtained by crossing the initial-state (anti)quark with the final-state gluon.}
\label{fig:realsm}
\end{figure}

\begin{figure}[h!]
\includegraphics[scale=0.7]{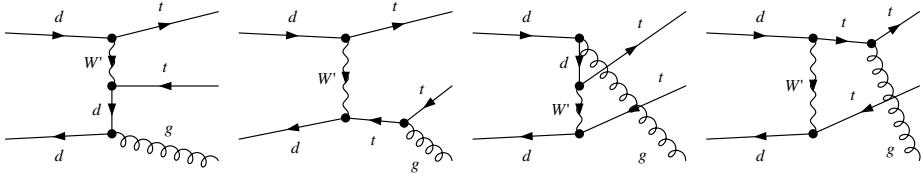}
\caption{Feynman diagrams for a gluon emission induced by NP interactions.
The diagrams for a (anti)quark emission can be obtained by crossing the initial-state (anti)quark with the final-state gluon.}
\label{fig:realnp}
\end{figure}

At the NLO level the real corrections consist of the radiations of an additional gluon or massless (anti)quark
in the final states as shown in Fig.\ref{fig:realsm} and Fig.\ref{fig:realnp}.

The phase space integration for the real gluon emission will produce soft and collinear
singularities, which can be isolated by slicing the phase space into different
regions using suitable cutoffs. In this paper, we use the two cutoff phase space slicing
method~\cite{PhysRevD.65.094032}, which introduces two arbitrary small cutoff parameters,
i.e. soft cutoff parameters $\delta_{s}$ and collinear parameters $\delta_{c}$, to decompose
the three-body phase space into three regions.

First, the phase space is separated into two regions by the soft cutoff parameters $\delta_{s}$ , according to
whether the energy of the emitted gluon is soft, i.e. $E_{5} \leqslant \delta_{s} \sqrt{s_{12}}/2 $, or hard,
i.e. $ E_{5} > \delta_{s} \sqrt{s_{12}} /2 $.
Then the collinear cutoff parameters $\delta_{c}$ is introduced to divide the hard gluon phase space into two regions,
according to whether the Mandelstam variables $t_{i5} \equiv (p_{i}- p_{5} )^{2}$ (i=1,2)
satisfy the collinear condition $  |t_{i5}| < \delta_{c}s_{12}$ or not.
Thus we  have
\begin{equation}
 d\hat{\sigma}^{Real} = d\hat{\sigma}^{S}+d\hat{\sigma}^{HC}+d\hat{\sigma}^{H\overline{C}}.
\end{equation}
The hard non-collinear term $d\hat{\sigma}^{H\bar{C}} $ can be written as,
\begin{equation}
d\hat{\sigma}^{H\overline{C}} =\frac{1}{2s_{12}} \int |\M_{3}|^2 d\Gamma_{3}|_{H\overline{C}}
\end{equation}
which can be evaluated numerically using standard Monte-Carlo techniques~\cite{Lepage:1977sw}.
In the following sections, we discuss the parts containing the soft and hard collinear sigularities.

In the limit that the energy of the emitted gluon becomes small, i.e. $E_{5} \leqslant \delta_{s} \sqrt{s_{12}}/2$,
the three-body cross section $d\hat{\sigma}^{S}$ can be factorized as
\begin{equation}\label{eq:soft}
 d\hat{\sigma}^{S}=  \left[\frac{\alpha_{s}}{2\pi} C'_{\epsilon} \right] \sum_{i,j=1}^{4}\left(d\hat{\sigma}^{\rm LO}_{\rm INT} \frac{\mathcal{C}_{ij}^{\rm INT}}{\mathcal{C}_{0}^{\rm INT}} +  d\hat{\sigma}^{\rm LO}_{\rm NPS} \frac{\mathcal{C}_{ij}^{\rm NPS}}{\mathcal{C}_{0}^{\rm NPS}}\right)\int dS \frac{- p_{i}\cdot p_{j}}{\left(p_{i}\cdot p_{5}\right)\left( p_{j}\cdot p_{5}\right)},
\end{equation}
where $C'_{\epsilon }=\frac{ \Gamma(1-\epsilon)}{ \Gamma(1-2\epsilon)}\left( \frac{4\pi \mu_{r}^2}{s_{12}}\right)^{\epsilon} $.
The color charge factors $\mathcal{C}_{ij}$ are
\begin{eqnarray}
\mathcal{C}_{12}^{\rm INT}&=&\mathcal{C}_{34}^{\rm INT}=C_{F}/2,~~~~~~~~~~\mathcal{C}_{33}^{\rm INT}=\mathcal{C}_{44}^{\rm INT}=C_{A}C_{F}^{2}, \nno\\
\mathcal{C}_{13}^{\rm INT}&=&\mathcal{C}_{24}^{\rm INT}=-C_{A}C_{F}^2,~~~~~~~\mathcal{C}_{14}^{\rm INT}=\mathcal{C}_{23}^{\rm INT}=-C_{F}/2 ,\nno\\
\mathcal{C}_{11}^{\rm INT}&=&\mathcal{C}_{22}^{\rm INT}=0,
\end{eqnarray}
and
\begin{eqnarray}
\mathcal{C}_{13}^{\rm NPS}&=&\mathcal{C}_{24}^{\rm NPS}=3C_{A}C_{F},~~~~~~~~~~\mathcal{C}_{33}^{\rm NPS}=\mathcal{C}_{44}^{\rm NPS}=-3C_{A}C_{F}, \nno\\
\mathcal{C}_{11}^{\rm NPS}&=&\mathcal{C}_{22}^{\rm NPS}=0, ~~~~~~~~~~~~\mathcal{C}_{12}^{\rm NPS}=\mathcal{C}_{34}^{\rm NPS}=\mathcal{C}_{14}^{\rm NPS}=\mathcal{C}_{23}^{\rm NPS}=0.
\end{eqnarray}
Here, $\mathcal{C}_{0}^{\rm INT}=C_{A}C_{F}$ and $\mathcal{C}_{0}^{\rm NPS}=9$  are the color factors of LO diagrams in Fig.~\ref{fig:FeynDiaLO}.

The integration over the soft phase space is given by~\cite{PhysRevD.65.094032}:
\begin{equation}
\int dS = \frac{1}{\pi} \left(\frac{4}{s_{12}}\right)^{-\epsilon} \int_{0}^{\delta_{s}\sqrt{s_{12}}/2} dE_{5} E^{1-2\epsilon}_{5} \sin^{1-2\epsilon}\theta_{1}d\theta_{1}\times \sin^{-2\epsilon}\theta_{2}d\theta_{2}.
\end{equation}
We define
\begin{equation}
\mathcal{I}_{ij}=  \int dS \frac{1}{\left(p_{i}\cdot p_{5}\right)\left( p_{j}\cdot p_{5}\right)}.
\end{equation}
Then we have
\begin{eqnarray}
\mathcal{I}_{11}&=& \mathcal{I}_{22}=0, \nno\\
\mathcal{I}_{33}&=& \mathcal{I}_{44}=-\frac{1}{ m_{t}^2}\frac{1}{\epsilon_{\rm IR}}+ \mathcal{I}_{33}^{fin}, \nno\\
 \mathcal{I}_{34}&=&\frac{2}{s}\left(-\frac{1}{\epsilon_{\rm IR}}\frac{1}{\beta}\ln \frac{\beta +1}{1-\beta }\right)+ \mathcal{I}_{34}^{fin}, \nno\\
 \mathcal{I}_{12} &=& \frac{2}{s} \left\lbrace \frac{1}{\epsilon_{\rm IR}^2}+\frac{1} {\epsilon_{\rm IR}} \left( -2 \ln \delta_s \right) \right\rbrace+ \mathcal{I}_{12}^{fin}, \nno\\
 \mathcal{I}_{13}&=&\mathcal{I}_{24} = \frac{1}{ t_{1}} \left\lbrace  -\frac{1}{\epsilon_{\rm IR}^{2}} + \frac{1}{\epsilon_{\rm IR}} \left(2 \ln \frac{-t_{1}}{s} +\ln \frac{s}{m_{t}^{2}}+2 \ln \delta_s \right) \right\rbrace+ \mathcal{I}_{13}^{fin}, \nno\\
 \mathcal{I}_{14}&=&\mathcal{I}_{23}= \frac{1}{ u_{1}} \left\lbrace  -\frac{1}{\epsilon_{\rm IR}^{2}} + \frac{1}{\epsilon_{\rm IR}} \left(2 \ln \frac{-u_{1}}{s} +\ln \frac{s}{m_{t}^{2}}+2 \ln \delta_s \right) \right\rbrace + \mathcal{I}_{14}^{fin},
  \end{eqnarray}
where all the IR sigularities in $\mathcal{I}_{ij}$ have been extracted out and for briefness, the finite terms $\mathcal{I}_{ij}^{fin}$ are not shown here.
Now, the Eq.(\ref{eq:soft}) can be rewritten as
 \begin{eqnarray}\label{eqs}
 d \hat{\sigma}^{S}&=&  \frac{\alpha _{s}}{2 \pi} C_{\epsilon} \left[\frac{(A_{2}^{S})_{\rm INT}}{\epsilon_{\rm IR} ^{2}}+\frac{ (A_{1}^{S})_{int} }{\epsilon_{\rm IR}}+(A_{0}^{S})_{\rm INT}\right]  d\hat{\sigma}^{\rm LO}_{\rm INT}, \nno\\
 &&+ \frac{\alpha _{s}}{2 \pi} C_{\epsilon} \left[\frac{(A_{2}^{S})_{\rm NPS}}{\epsilon_{\rm IR} ^{2}}+\frac{(A_{1}^{S})_{\rm NPS}}{\epsilon_{\rm IR}}+(A_{0}^{S})_{\rm NPS}\right] d\hat{\sigma}^{\rm LO}_{\rm NPS},
\end{eqnarray}
in which
\begin{eqnarray}\label{eqasint}
(A^{S}_{2})_{\rm INT}&=& 2C_{F},\nno \\
(A^{S}_{1})_{\rm INT}&=& -\frac{1} {C_{A}} \left[  16 \ln \frac{-t_{1}} {\mu_{r} ^{2}}+2 \ln \frac{-u_{1}} {\mu_{r} ^{2}}   +9 \ln \frac{\mu_{r} ^{2}} {m_{t}^{2}}
  + \ln \frac{\mu_{r} ^{2}} {s} \right.\nno \\&&
  \left.+16\ln \delta_{s}- \frac{ \left(1+\beta ^{2}\right)} {2 \beta} \ln \frac{1+\beta} {1-\beta} -8 \right],
\end{eqnarray}
and
\begin{eqnarray}\label{eqasnps}
(A^{S}_{2})_{\rm NPS}&=& 2C_{F}, \nno\\
(A^{S}_{1})_{\rm NPS}&=& -2C_{F} \left[  2 \ln \frac{-t_{1}} {\mu_{r} ^{2}} + \ln \frac{\mu_{r} ^{2}} {m_{t}^{2}} +2\ln \delta _{s} -1  \right].
\end{eqnarray}

In the hard collinear region, $E_{5} > \delta_{s} \sqrt{s_{12}} /2 $  and $ |t_{i5}| < \delta_{c} s_{12}$,
the emitted hard gluon is collinear to one of the incoming partons and
the three-body cross section is factorized as
\begin{eqnarray}
d\sigma^{HC}&=&d\sigma^{\rm LO}\left[\frac{\alpha_{s}}{2\pi}C'_{\epsilon}\right]\left(-\frac{1}{\epsilon}\right)\delta_{c}^{-\epsilon}\left[P_{dd}(z,\epsilon)G_{d/p}(x_{1}/z)G_{\bar{d}/p}(x_2)\right. \nno\\
&&\left.+P_{\bar{d}\bar{d}}(z,\epsilon)G_{\bar{d}/p}(x_{1}/z)G_{d/p}(x_2)+(x_1\leftrightarrow x_2)\right]\frac{dz}{z} \left(\frac{1-z}{z}\right)^{-\epsilon}dx_1 dx_2,
\end{eqnarray}
where $P_{ij}$ are the unregulated splitting functions in $n = 4-2\epsilon$ dimension for $0 < z < 1$,
which is related to the usual Altarelli-Parisi splitting kernels~\cite{Altarelli:1977zs}
as $P_{ij} (z, \epsilon) = P_{ij} (z) + \epsilon P'_{ij} (z)$. Explicitely, in our case,
\begin{eqnarray}
P_{dd}(z)&=& P_{\bar{d}\bar{d}}(z)=C_{F}\frac{1+z^2}{1-z},\\
P'_{dd}(z)&=& P'_{\bar{d}\bar{d}}(z)=-C_{F}(1-z).
\end{eqnarray}

For massless $d(\bar{d})$ emission, we decompose the phase space into two regions, collinear and non-collinear,
and give the expression for $gd \rightarrow t\bar{t} d $ cross section,
\begin{eqnarray}
d\sigma(qg \rightarrow t\bar{t} q)&=&\sum_{\alpha =d,\bar{d}} \hat{\sigma}^{\overline C}(\alpha g \rightarrow t\bar{t} \alpha)[G_{\alpha /p}(x_1)G_{g/p}(x_2)+(x_1\leftrightarrow x_2)]dx_1 dx_2 \nno\\
&+& d\sigma^{\rm LO} \left[\frac{\alpha_{s}}{2\pi}C'_{\epsilon}\right]\left(-\frac{1}{\epsilon}\right)\delta_{c}^{-\epsilon}[P_{dg}(z,\epsilon)G_{g/p}(x_{1}/z)G_{\bar{d}/p}(x_2) \nno \\
&+& P_{\bar{d}g}(z,\epsilon)G_{g/p}(x_{1}/z)G_{d/p}(x_2)+(x_1\leftrightarrow x_2)]\frac{dz}{z} \left(\frac{1-z}{z}\right)^{-\epsilon}dx_1 dx_2,
\end{eqnarray}
where
\begin{eqnarray}
 P_{dg}(z)=P_{\bar d g}(z)=\frac{1}{2}[z^2+(1-z)^2], \qquad   P'_{dg}(z)=P'_{{\bar d} g}(z)=-z(1-z),
\end{eqnarray}
and
\begin{equation}
\hat{\sigma}^{\overline C}(\alpha g \rightarrow t\bar{t} \alpha) =\frac{1}{2s_{12}} \int |\M_{3}|^2(\alpha g \rightarrow t\bar{t} \alpha) d{\Gamma_{3}}|_{H\overline{C}}.
\end{equation}

In order to factorize the collinear singularity into the PDF,
we introduce a scale dependent PDF in the  $\overline{\rm MS}$ convention~\cite{PhysRevD.65.094032},
\begin{equation}
G_{\alpha/p}(x,\mu_{f})=G_{\alpha/p}(x)+\sum_{\beta} \left(-\frac{1}{\epsilon}\right)\left[\frac{\alpha_{s}}{2 \pi} \left(\frac{4\pi \mu_{f}^2}{\mu_{r}^2}\right)^{\epsilon}\right]\int_{x}^{1}\frac{dz}{z} P_{\alpha \beta}(z)G_{\beta /p}(x/z).
\end{equation}
As in Ref.~\cite{PhysRevD.65.094032}, the  $\mathcal{O}(\alpha_{s})$ collinear contribution is
\begin{eqnarray}\label{eqc}
d\sigma^{coll}&=&d\hat{\sigma}^{\rm LO}\left[\frac{\alpha_{s}}{2\pi}C'_{\epsilon}\right] \big\lbrace\tilde{G}_{d/p}(x_1,\mu_{f})G_{\bar d/p}(x_2,\mu_{f})+G_{d/p}(x_1,\mu_{f})\tilde{G}_{\bar d/p}(x_2,\mu_{f}) \nno \\
&+&\sum_{\alpha= d,\bar d}\left[ \frac{A_{1}^{sc}(\alpha \rightarrow \alpha g)}{\epsilon}+A_{0}^{sc}(\alpha \rightarrow \alpha g)\right]G_{d/p}(x_1,\mu_{f})G_{\bar d/p}(x_2,\mu_{f}) \nno\\
&& +(x_1\leftrightarrow x_2)\big\rbrace dx_1dx_2,
\end{eqnarray}
 where
 \begin{eqnarray}\label{eqac}
     A_{1}^{sc}(d\rightarrow d g)&=& A_{1}^{sc}(\bar{d} \rightarrow \bar{d} g)=C_{F}(2\delta_s+3/2), \nno \\
     A_{0}^{sc}&=&A_{1}^{sc}\ln\frac{s_{12}}{\mu_{f}^2},
 \end{eqnarray}
 and
  \begin{eqnarray}
  G_{\alpha /p}(x_1,\mu_{f})&=&\sum_{\beta} \int _{x}^{1-\delta s\delta \alpha \beta } \frac{dy}{y} G_{\beta /p}(x_1,\mu_{f})\tilde{P}_{\alpha \beta}(y),
\end{eqnarray}
with
\begin{equation}
 \tilde{P}_{\alpha \beta}(y)=P_{\alpha \beta}(y)\ln\left(\delta c\frac{1-y}{y}\frac{s_{12}}{\mu_{f}^2}\right) - P'_{\alpha \beta}(y).
\end{equation}

Finally the NLO correction of $d \bar{d} \to t\bar{t}$ process can be written as
\begin{eqnarray}\label{eqtot}
\sigma^{\rm NLO} &=& \int \left\lbrace  d x_{1} d x_{2} \left[G_{d/p}(x_{1},\mu_{f})G_{\bar{d}/p}(x_{2},\mu_{f})+(x_{1} \leftrightarrow x_{2})\right] (  \sigma^{virt}+\sigma^{S}+\sigma^{H\bar{C}})+\sigma^{coll}\right\rbrace \nno \\
&+& \sum_{\alpha =d,\bar{d}} \int d x_{1} d x_{2} \left[G_{g/p}(x_{1},\mu_{f})G_{\alpha /p}(x_{2},\mu_{f})+(x_{1}\leftrightarrow x_{2})\right] \sigma ^{\overline{C}} (g\alpha \ra t\bar{t}\alpha).
\end{eqnarray}

Note that all the IR divergences in the NLO total cross section are proportional to the LO cross sections.
and we find the following relations
\begin{eqnarray}
 (A_{2}^{v})_{\rm INT}+(A_{2}^{S})_{\rm INT}=0,~~~~(A_{1}^{v})_{\rm INT}+( A_{1}^{S})_{\rm INT}+ \sum_{\alpha=d, \bar{d}}A_{1}^{sc}(\alpha \rightarrow \alpha g)=0, \nno\\
 (A_{2}^{v})_{\rm NPS}+(A_{2}^{S})_{\rm NPS}=0,~~~~(A_{1}^{v})_{\rm NPS}+( A_{1}^{S})_{\rm NPS}+ \sum_{\alpha=d, \bar{d}}A_{1}^{sc}(\alpha \rightarrow \alpha g)=0.
 \end{eqnarray}
Now all the IR divergences are canceled exactly.

\section{Numerical Results}\label{s4}
In the numerical calculations, we set $m_{ W'}= 400 $ GeV,
because such a $W'$ is readily observed at Tevatron with an integrated luminosity of $10 ~\rm fb ^{-1}$,
and at the LHC with an integrated luminosity of $100 ~\rm pb^{-1}$~\cite{Cheung:2011qa}.
There are two independent parameters in the NP Lagrangian.
For the convenience of calculations we define the a parameter set $(\rm C_{\rm {V}}, \rm C_{\rm A})$,
where $ \rm {C}_{\rm {V}}= g'(f_{R}+f_{L})/2$ and $\rm C_{\rm {A}}=g'(f_{R}-f_{L})/2 $.
The mass of top quark is chosen to be $m_t=172.5~{\rm GeV}.$
The CTEQ6L and CTEQ6M PDF sets~\cite{Pumplin:2002vw} and the associated $\as$ functions are used
for LO and NLO calculation, respectively. Both the renormalization and factorization scales are fixed to the top quark mass unless
specified otherwise.

\begin{figure}[htbp]
\centerline{\hbox{
\includegraphics[width=0.7\textwidth]{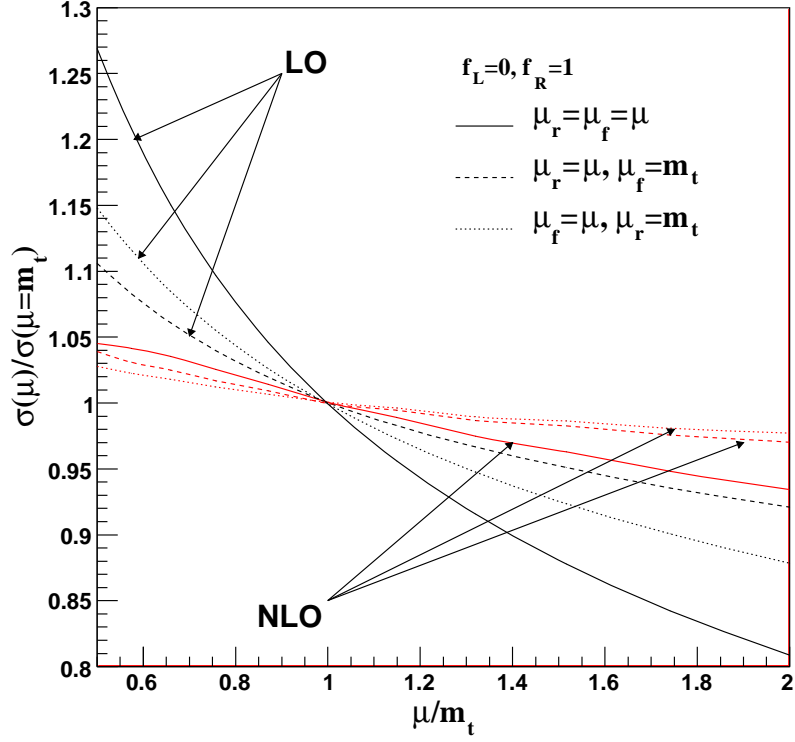}}}
 \caption{\label{fig:scale_dep} Scale dependences of the total cross
sections at the Tevetron. The black and the red lines represent the LO
and NLO results, respectively.}
\end{figure}
\subsection{scale dependence}
In Fig.~\ref{fig:scale_dep} we show the scale dependence of the LO
and NLO total cross sections at the Tevatron for three cases: (1) the
renormalization scale dependence $\mu_r=\mu,\ \mu_f=m_t$, (2) the
factorization scale dependence $\mu_r=m_t,\ \mu_f=\mu$, and (3)
total scale dependence $\mu_r=\mu_f=\mu$.
From Fig.~\ref{fig:scale_dep}, we can see that
the NLO corrections significantly reduce the scale dependence for all
three cases, making the theoretical predictions more reliable.

\begin{figure}[h!]
\begin{centering}
\includegraphics[scale=0.5]{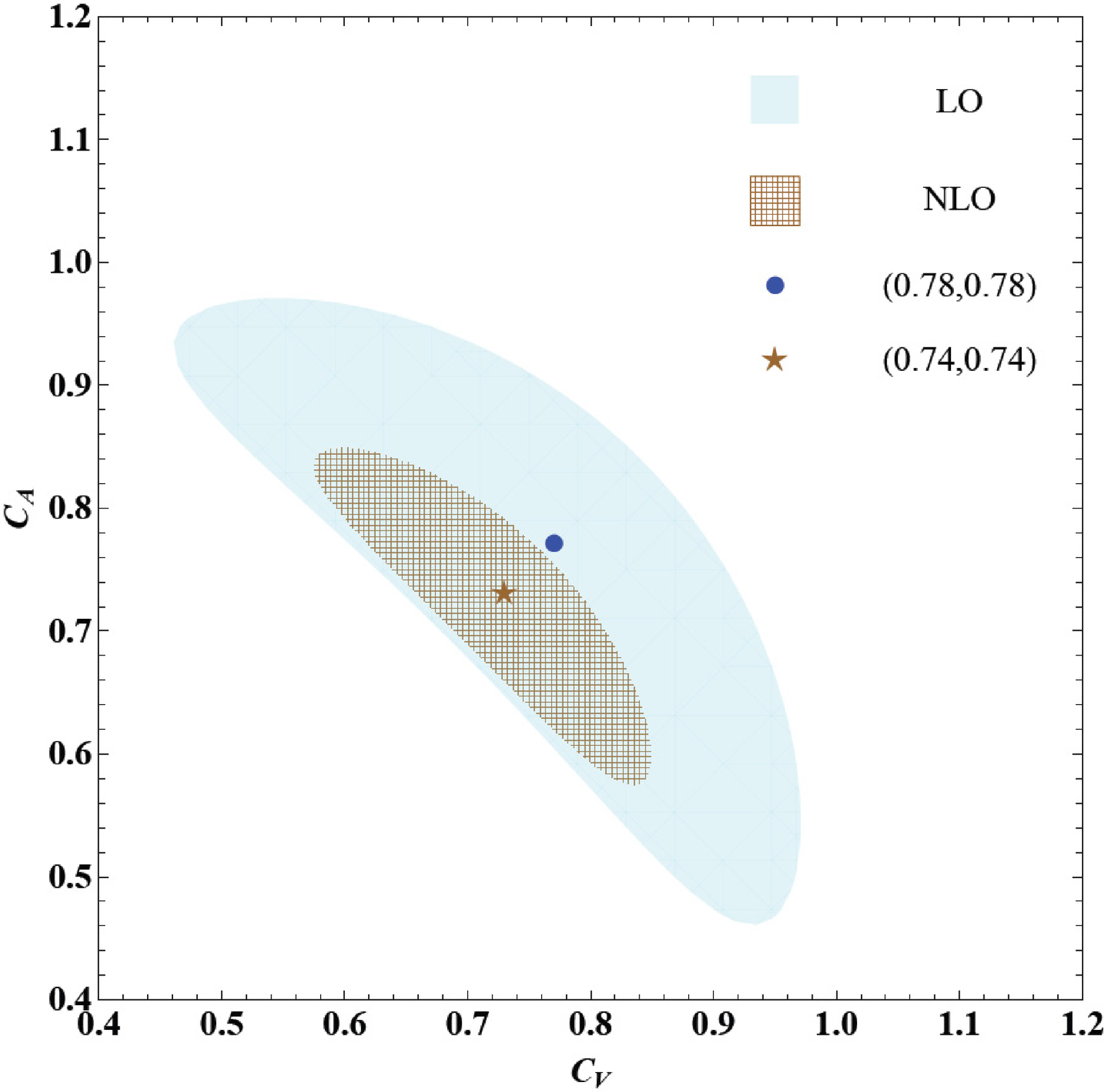}
\end{centering}
\caption{\label{fig:octet_para_scale} Values of $\rm{C}_{\rm{V}}$ and
$\rm{C}_{\rm{A}}$ allowed by  Tavetron data at 95\% CL:
$\sigma_{t\bar t}$=(7.50 $\pm$ 0.48)~pb ,
$A_{\textrm{FB}}(m_{t \bar t} > 450 ~\rm{ GeV})$=0.475$\pm$0.114,  and
$\left(d\sigma_{t \bar{t}} /d m_{t\bar{t}}\right)^{m_{t \bar{t}} \in [800,1400] ~\textrm{GeV} } = (0.068 \pm 0.036 ) ~\rm fb/GeV$. The blue dot (0.78, 0.78) and brown star (0.74, 0.74) represent the BFPs at LO and NLO level, respectively.
The allowed parameter region is symmetric with respect to the $\rm{C}_{\rm{A}} $  and $\rm{C}_{\rm{V}}$ axes, so we only display the contours where $\rm{C}_{\rm{A}} \rm{C}_{\rm{V}} >0 $. }
\end{figure}


\subsection{Tevatron constraints}

$A_{\textrm{FB}}$ of top quark pair productions is defined as
\begin{eqnarray}
  A_{\textrm{FB}}&=&\frac{\sigma_\textrm{F} - \sigma_\textrm{B}}{\sigma_\textrm{F} +
  \sigma_\textrm{B}}\nno\\
  &=&A_{\textrm{FB}}^{\textrm{NP}} \times R +
  A_{\textrm{FB}}^{\textrm{SM}}\times( 1 - R )\nno
\end{eqnarray}
where
\begin{eqnarray}
  A_{\textrm{FB}}^{\textrm{NP}}&=&(\sigma_\textrm{F}^{\textrm{NP}} -
  \sigma_\textrm{B}^{\textrm{NP}})/(\sigma_\textrm{F}^{\textrm{NP}} +
  \sigma_\textrm{B}^{\textrm{NP}}),\nno\\
  A_{\textrm{FB}}^{\textrm{SM}}&=&(\sigma_\textrm{F}^{\textrm{SM}} -
  \sigma_\textrm{B}^{\textrm{SM}})/(\sigma_\textrm{F}^{\textrm{SM}} +
  \sigma_\textrm{B}^{\textrm{SM}}),\nno\\
  R&=&\sigma_{\textrm{tot}}^{\textrm{NP}}/(\sigma_{\textrm{tot}}^{\textrm{SM}} + \sigma_{\textrm{tot}}^{\textrm{NP}})
\end{eqnarray}
are the asymmetries induced by NP and SM, and $R$ is the fraction of
NP contribution to the total cross section. $\sigma_{\textrm{F}}$
and $\sigma_{\rm B} $ denote the total cross sections in the
forward(F) and backward(B) rapidity regions, respectively.
The LO and NLO total cross sections of the interference  and NP contributions can be
written in terms of $\textrm{C}_{\textrm{V}}^{2}$ and $\textrm{C}_{\textrm{A}}^{2}$,
\begin{eqnarray}\label{paracslo}
  [\sigma^{\textrm{INT}}_{t\bar t}]_{\textrm{LO}} &=& \left[-(1.14)^{+0.22}_{-0.31}(\textrm{C}_{\textrm{V}}^{2}+\textrm{C}_{\textrm{A}}^2)
 \right]~\textrm{pb},\\
 \left[ \sigma^{\textrm{NPS}}_{t\bar t}\right] _{\textrm{LO}} &=& \left[  2.06_{-0.27}^{+0.35}(\textrm{C}_{\textrm{V}}^{2}+\textrm{C}_{\textrm{A}}^2)^{2} - (2.51)^{+0.32}_{-0.40}(\textrm{C}_{\textrm{V}}^{2}  \cdot\textrm{C}_{\textrm{A}}^2) \right]~\textrm{pb},
 \end{eqnarray}
and
\begin{eqnarray}
 [ \sigma^{\textrm{INT}}_{t\bar t}]_{\textrm{NLO}} &=& \left[-(1.42)_{-0.10}^{+0.06}) (\textrm{C}_{\textrm{V}}^{2}+\textrm{C}_{\textrm{A}}^{2}) \right] ~\textrm{pb},\\
  \left[ \sigma^{\textrm{NPS}}_{t\bar t}\right] _{\textrm{NLO}}&=& \left[ 2.39_{-0.09}^{+0.06}(\textrm{C}_{\textrm{V}}^{2}+\textrm{C}_{\textrm{A}}^2)^{2} - \left(2.82\right)_{-0.10}^{+0.04}(\textrm{C}_{\textrm{V}}^{2}  \cdot \textrm{C}_{\textrm{A}}^2) \right]~\textrm{pb},
\end{eqnarray}
where the errors are obtained by varying the scale between $\mu_r =\mu_f = m_t/2$ and $\mu_r = \mu_f = 2m_t$.
The differences of the cross sections in the forward and backward
rapidity region are given by
\begin{eqnarray}\label{paraafblo}
  \left[ \sigma_{\textrm{F}}^{\textrm{INT}} -
  \sigma_{\textrm{B}}^{\textrm{INT}}\right] _{\textrm{LO}} &=&
 \left[-(0.26)_{+0.08}^{-0.05}(\textrm{C}_{\textrm{V}}^{2}+\textrm{C}_{\textrm{A}}^{2})\right] ~\textrm{pb},\\
  \left[\sigma_{\textrm{F}}^{\textrm{NPS}}-
   \sigma_{\textrm{B}}^{\textrm{NPS}}\right] _{\textrm{LO}} &=&
  \left[ 0.68^{-0.10} _{+0.12}(\textrm{C}_{\textrm{V}}^{2}+\textrm{C}_{\textrm{A}}^{2})^{2} +0.054^{-0.007}_{+0.009}(\textrm{C}_{\textrm{V}}^{2} \cdot \textrm{C}_{\textrm{A}}^{2})\right]~ \textrm{pb},
\end{eqnarray}
and
\begin{eqnarray}\label{paraafbnlo}
  \left[ \sigma_{\textrm{F}}^{\textrm{INT}} -
  \sigma_{\textrm{B}}^{\textrm{INT}} \right] _{\textrm{NLO}} &=&
 \left[ -(0.40)_{+0.04}^{-0.05}(\textrm{C}_{\textrm{V}}^2+\textrm{C}_{\textrm{A}}^2)\right] ~\textrm{pb}, \\
\left[  \sigma_{\textrm{F}}^{\textrm{NPS}} - \sigma_{\textrm{B}}^{\textrm{NPS}}\right] _{\textrm{NLO}} &=&
  \left[ 0.94^{-0.03} _{+0.04}(\textrm{C}_{\textrm{V}}^2+\textrm{C}_{\textrm{A}}^2)^2 -(0.127)^{-0.000}_{+0.002} (\textrm{C}_{\textrm{V}}^2 \cdot \textrm{C}_{\textrm{A}}^2)\right]~ \textrm{pb}.
\end{eqnarray}
For $t\bar{t} $ invariant mass spectrum, we restrict our attention in the large invariant mass region,
i.e. $m_{t\bar t} \in [800,1400] ~\textrm{GeV}$, where the $\rm A_{\rm FB}$ is the most obvious. The results are presented as
\begin{eqnarray}
\left[\frac{d\sigma^{\rm INT}_{t \bar {t}}} {d m_{t\bar t}} \right]_{\textrm{LO}} ^{m_{t\bar t} \in [800,1400]} &=&\left[ -(0.014)^{+0.006}_{-0.004}(\textrm{C}_{\textrm{V}}^2+\textrm{C}_{\textrm{A}}^2)\right] ~\frac{\rm pb}{\rm GeV}, \\
\left[\frac{d\sigma^{\rm NPS}_{t \bar {t}}} {d m_{t\bar t}} \right]_{\textrm{LO}} ^{m_{t\bar t} \in [800,1400]}&= & \left[0.082^{+0.020}_{-0.018}(\textrm{C}_{\textrm{V}}^2+\textrm{C}_{\textrm{A}}^2)^2- (0.064)^{+0.007}_{-0.008}(\textrm{C}_{\textrm{V}}^2 \cdot \textrm{C}_{\textrm{A}}^2)\right]~  \frac{\rm pb}{\rm GeV},
\end{eqnarray}
and
\begin{eqnarray}
\left[\frac{d\sigma^{\rm INT}_{t \bar {t}}} {d m_{t\bar t}}\right]_{\textrm{NLO}} ^{m_{t\bar t} \in [800,1400]}  &=& \left[-(0.012)^{+0.004}_{-0.002}(\textrm{C}_{\textrm{V}}^2+\textrm{C}_{\textrm{A}}^2) \right] ~\frac{\rm pb}{\rm GeV},\\
 \left[\frac{d\sigma^{\rm NPS}_{t \bar {t}}} {d m_{t\bar t}}\right]_{\textrm{NLO}} ^{m_{t\bar t} \in [800,1400]}  &=& \left[0.117^{+0.014}_{-0.010}(\textrm{C}_{\textrm{V}}^2+\textrm{C}_{\textrm{A}}^2)^2- (0.094)^{+0.003}_{-0.004}(\textrm{C}_{\textrm{V}}^2 \cdot \textrm{C}_{\textrm{A}}^2)\right] ~ \frac{\rm pb}{\rm GeV},
 \label{eq:invm}
\end{eqnarray}
 From the errors in Eqs.(\ref{paracslo} - \ref{eq:invm}) we can see that NLO corrections reduce the dependence
 of the cross sections on the renormalization and factorization scales.

In Fig.~\ref{fig:octet_para_scale}, we show the allowed
region in the $(\rm C_{\rm V},\rm C_{\rm A})$ plane that
is consistent with the total cross section $\sigma_{t\bar t}$, $\rm A_{\rm FB}$~\cite{Aaltonen:2011kc}
and the spectrum of $m_{t\bar t}$ in the large mass region~\cite{Aaltonen:2009iz}, which are given by
\begin{eqnarray}
\sigma_{t\bar{t}}^{\rm EX}&=&(7.50 \pm 0.48)\textrm{pb}, \\
\rm{A_{FB}^{EX}}&=&0.475 \pm 0.114,~~~~~~~\rm for\textrm{ m}_{t\bar{t}} > 450 \textrm{GeV}, \\
\left[\frac{d\sigma_{t \bar t}}{d m_{t\bar{t}}}\right]^{\textrm{m}_{t \bar{t}} \in [800,1400] \textrm{GeV} }_{\rm {EX}} &=& (0.068 \pm 0.036 ) ~\rm fb/GeV.
\end{eqnarray}
We use Monte Carlo programm MCFM~\cite{Campbell:1999ah} to get the cross section of the gluon fusion
channel $gg \rightarrow t\bar t$ at the NLO QCD level. As for the process of $q\bar q \rightarrow t\bar t$, we have checked our
value with the results given by MCFM at QCD NLO level, which are well consistent in the
range of Monte Carlo integration error. Combining the contributions of these two channels
we have the SM predictions for the above observables at  NLO level
\begin{eqnarray}
\sigma_{t\bar{t}}^{\textrm{SM}}&=&7.00^{+0.36}_{-0.76}~\textrm{pb}, \\
\left[\frac{d\sigma_{t \bar t}}{d m_{t\bar{t}}}\right]^{\textrm{m}_{t \bar{t}} \in [800,1400] \textrm{GeV} }_{\rm {SM}} &=& 0.055^{ +0.010}_{-0.005}  ~\rm fb/GeV,
\end{eqnarray}
where we have considered scale uncertainty in the calculations. For consistency we have used the SM QCD
predicted values of $\rm A_{FB}\left( m_{t\bar t} \geq 450
\rm{GeV} \right)= 0.088 \pm 0.013$ at NLO level, although
next-to-next-to-leading logarithmic (NNLL) SM QCD results are available~\cite{Ahrens:2011uf}.\\
\indent The measurements of $A_{\rm FB}$
and invariant mass spectrum $d\sigma_{t\bar{t}}/ d m_{t\bar{t}}$  in the large invariant mass region are particularly sensitive to  values of $\rm C_{\rm V}$ and $\rm C_{\rm A} $  at NLO level. In order to generate the desired $\rm A_{\rm FB}$  in large  $m_{t \bar t} $ region, NP couplings should be large enough so that the positive NPS terms could overcome the negative INT terms. While on the other hand, the NLO NPS effect causes the cross section in the last bin of $m_{t\bar t}$  to exceed the $1 \sigma$ upper limit of experimental  result and therefore we expect the couplings $ \rm C_{\rm V}$ and $ \rm C_ {\rm A }$ to be not too large.  As a consequence,  NP couplings are subject to strong restrictions.

In Figs.~\ref{fig:octet_para_scale}, solid and grid regions correspond
to NP LO and NLO results at $95\%  $ confidence level~(CL), where we
have considered theoretical and experimental uncertainty in
$ \sigma_{t \bar{t}}$ and  $ d\sigma_{t \bar t} / \textrm{m}_{t \bar t} $ and only consider experimental uncertainty in
the $\textrm{A}_{\textrm{FB}}$ calculation. It can be seen that NLO
corrections manifestly trim down  the  area of the allowed parameter region.  The
blue dot $\rm (0.78,0.78)$ and brown star $\rm
(0.74,0.74)$ represent the best fit points (BFPs) at LO
and NLO level, where $ \chi  ^{2}$ reaches its minimums of 2.0 and 2.3, respectively.
Thus we can see that higher order
corrections impose stronger constraints on NP couplings and  reduce the BFPs of $\textrm{C}_{\textrm{V}}$ and $\textrm{C}_{\textrm{A}}$ by 5\%.

 Now we discuss the  theoretical predictions for the measurements at Tevatron induced by NP at the NLO BFP $\rm (0.74,0.74)$,  or equivalently,
$ g'=1.48,$ and $f_{R}=1,~ f_{L}=0 $ :
The LO and NLO total cross sections of $t\bar t$ production are
\begin{eqnarray}
  \sigma ^{\rm NP} _{t\bar {t},\rm LO}= 0.461 ~\rm{pb} ,\nno\\
  \sigma ^{\rm NP}_{t\bar {t},\rm NLO}= 0.458 ~\rm{pb},
\end{eqnarray}
and the differential cross section are
\begin{eqnarray}
\left[d\sigma^{\rm NP}_{t \bar {t}} / d  m_{t \bar t} \right]^{\rm m _{t \bar t} \in [800, 1400] \rm {GeV}}_{\rm LO}&=& 0.062 ~\rm fb/GeV, \nno\\
\left[d\sigma^{\rm NP}_{t \bar {t}} /d  m_{t \bar t} \right]^{\rm m_{t \bar t} \in [800, 1400] \rm {GeV}}_{\rm NLO}&=& 0.083 ~\rm fb/GeV.
\end{eqnarray}
Here, the superscript "NP" represents the combination of the "INT" and "NPS" contributions mentioned above.
It can be seen that the NLO corrections have slight effects on the total cross section
but increase the invariant mass distribution in the large mass region.
Note that the two parts of the NP corrections, INT and NPS terms, are individually not small, but
they have opposite sign and cancel each other.

The $\rm A_{\rm FB}$ containing NP contributions at the NLO BFP are shown in Table~\ref{tab:bestfit}.
All the theoretical predictions containing NP NLO effects are consistent with experimental results within 2$\sigma$ CL.
\begin{table*}[h!]
\begin{center}
\scalebox{1}[0.9] {\begin{tabular}{c|c|c}
  \hline
  \hline
  &SM NLO QCD + NP LO &SM NLO QCD + NP NLO  \\
  \hline
  $A_{\rm FB}^{p\bar p}$  &0.130  & 0.140 (0.2 $\sigma$) \\
  $A_{\rm FB}^{t\bar t}$  & 0.146  & 0.156 (0.0 $\sigma$) \\
  $A_{\rm FB}^{t\bar t}\left(m_{t\bar t} < 450~\rm GeV\right)$  & 0.074  & 0.068 (1.2 $\sigma$) \\
  $A_{\rm FB}^{t\bar t}\left(m_{t\bar t} > 450~\rm GeV\right)$  & 0.250 & 0.272 (1.8 $\sigma$) \\
  \hline
  \hline
\end{tabular}}
\end{center}
\caption{\label{tab:bestfit} The $\rm A_{FB}$ with $\rm g'=1.48$ ,  $f_{R} =1$, $f_{L}=0$  and $M_{W'}=400 ~\rm GeV$ at the Tevatron, where $A_{\rm FB}^{p\bar p}$ and $A_{\rm FB}^{t\bar t}$ are the $A_{\rm FB}$ in the lab frame and the $t\bar t$ rest frame, respectively. Here we list the CL when including NP effects at NLO level.}
\end{table*}
It is found that the $\rm{A}_{\rm{FB}}$ in the large invariant mass region gets an obvious enhancement by about 9\%.
\begin{figure}[h!]
\begin{centering}
\includegraphics[scale=0.65]{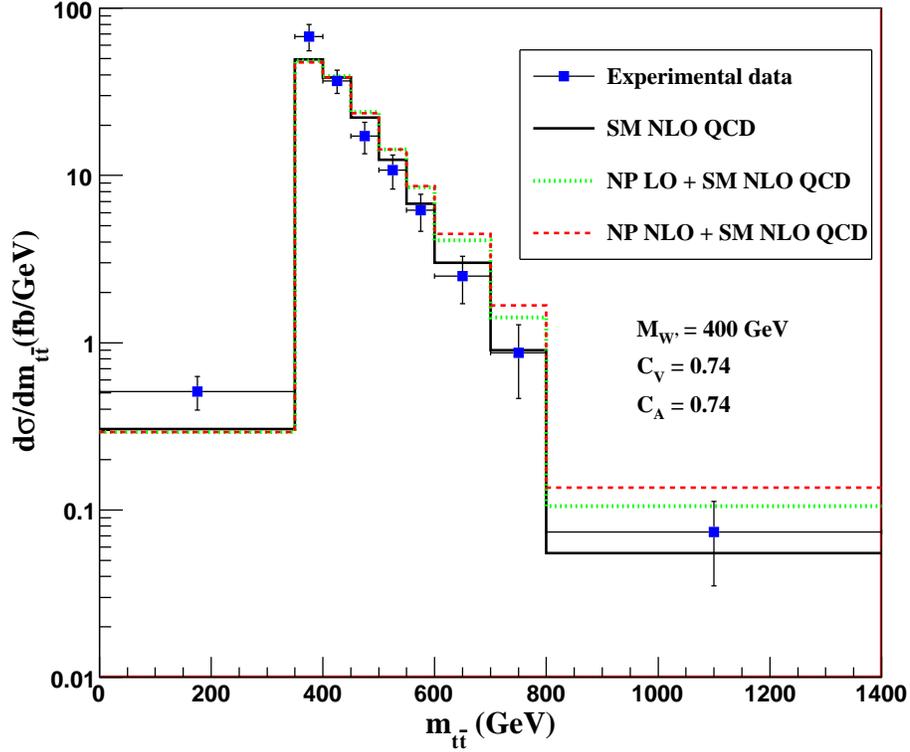}
\par\end{centering}
\caption{Differential cross sections $d\sigma/d
m_{t\bar{t}}$  as a function of  $m_{t\bar t}$ at the NLO BFPs $\rm (\pm 0.74,\pm 0.74)$. Here "Experimental data" is $d\sigma/d
m_{t\bar{t}}$ measured with 2.7 fb$^{-1}$ of integrated luminosity at the Tevatron~\cite{Aaltonen:2009iz}. "SM NLO QCD" represents the results in the SM QCD at NLO level. "NP LO + SM NLO QCD" and "NP NLO + SM NLO QCD" stand for the predictions including NP effects at LO and NLO level, respectively. }
\label{fig:mtt_tev}
\end{figure}

In Fig.~\ref{fig:mtt_tev}, we show differential cross section $d\sigma/d m_{t\bar{t}}$ when we consider NP effects at the
NLO BFP, from which we can see that higher order corrections do not change the distribution very much.

\begin{figure}[h!]
\begin{center}
\includegraphics[scale=0.7]{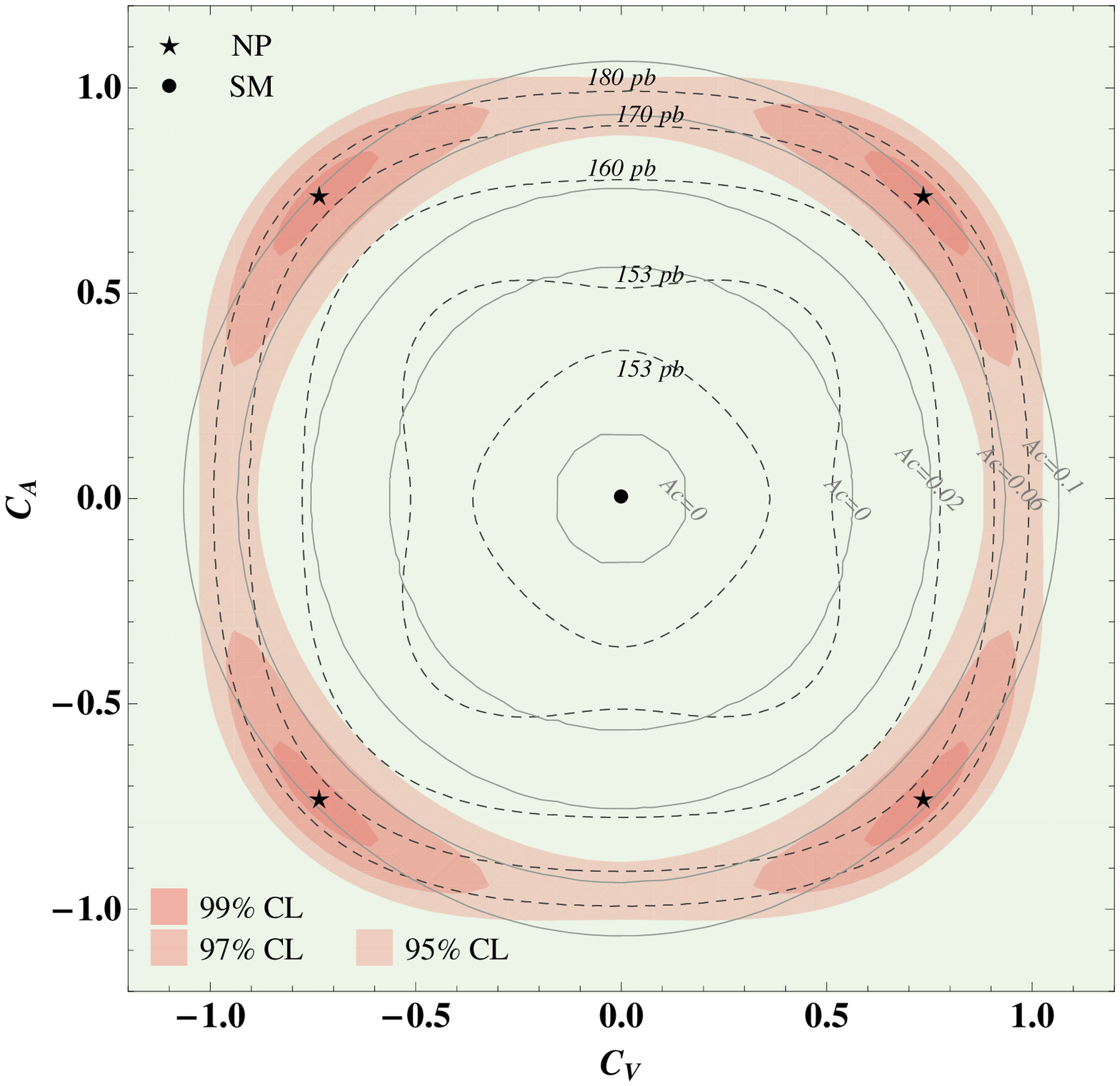}
\end{center} 
\caption{ Results of a combined fit to
$\sigma_{t\bar t}$ and the value of $\rm A_{FB}$ allowing for
NP at different CL. The shadows from dark to light indicate,
the experimentally favored region of 95\%, 97\%, and 99\%
probability in the $\rm C_{\rm V} - \rm C_{\rm A}$ plane.
The black dashed lines and grey solid lines, respectively,
represent the value of the total $t\bar t$ cross section and 
the $\rm A _{\rm C}$ at the LHC with $\sqrt{S}=7$ TeV. 
The black dot and stars represent the SM point and the NP NLO BFPs .
}
\label{fig:para_tota}
\end{figure}

\subsection{LHC predictions}
The process of top quark pair production has been measured at the LHC, and the cross
section~\cite{Roe:1392069,CMS-PAS-TOP-11-001} is:
\begin{eqnarray}
  \sigma_{t\bar t}^{\rm ATLAS}({\rm LHC, \sqrt{S}=7~TeV}) &=& 180 \pm 18 ~{\rm pb},\\
   \sigma_{t\bar t}^{\rm CMS}({\rm LHC, \sqrt{S}=7~TeV}) &=& 158 \pm 19 ~{\rm pb},
\end{eqnarray}
which are consistent with the SM predictions.
Since the LHC is a proton-proton collider, which is forward-backward symmetric, the $A_{FB}$
defined at Tevatron can not be directly applied to the proton-proton collider experiments
at the LHC. $\rm A_{\rm {C}}$ used by CMS ~\cite{CMS-PAS-TOP-10-010,CMS-PAS-TOP-11-014} is defined as
\begin{eqnarray}
\rm A_{\rm {C}} =\frac{\sigma(|\eta_{t}|-|\eta_{\bar{t}}|>0)-\sigma(|\eta_{t}|-|\eta_{\bar{t}}|<0)}{\sigma(|\eta_{t}|-|\eta_{\bar{t}}|>0)+\sigma(|\eta_{t}|-|\eta_{\bar{t}}|<0)}
\end{eqnarray}
where $\eta_{t}$ and $ \eta_{\bar{t}} $ are pseudo rapidities of top and antitop quark, respectively.
Last year at CMS, it is measured to be~\cite{CMS-PAS-TOP-10-010},  \[\rm A_{\rm {C}}  = 0.060 \pm 0.134(stat.) \pm 0.026(syst.)\]
whereas the recently updated report shows~\cite{CMS-PAS-TOP-11-014},
 \[\rm A_{\rm {C}}  = -0.016 \pm 0.030(stat.) \pm 0.019(syst.)\]
 The discrepancy between these two mesurements is evident. However, given the large experimental error, both results are compatible with the SM predictions $\rm A_{\rm {C}}  = 0.013 $~\cite{CMS-PAS-TOP-10-010}.  The $\rm A_{\rm C}$ induced
by NP interactions at the NP NLO BFPs $(\pm 0.74, \pm 0.74)$ is 0.069, which is about 5
times of SM prediction. This result is very close to the previous observed central value at CMS~\cite{CMS-PAS-TOP-10-010}, and is also consistent with the the latest data value~\cite{CMS-PAS-TOP-11-014} within 2 $\sigma$ CL.
 We still need more experimental data with higher precision  to seek evidence for a possible modification in  $\rm A_{\rm C}$ by NP.

In Fig.~\ref{fig:para_tota}, we show the results of a
combined fit to the $t\bar t$ data in the presence of NP at
different CL. The shadows from dark to light indicate the
experimentally preferred region of 95\%, 97\%, and 99\% probability
in the $\rm C_{\rm V}- \rm C_ {\rm A}$ plane. The black dot represents the SM point (0,0), and the black star represent the NLO BFP
$ (\rm C_{\rm V}, \rm C_ {\rm A})$. At the LHC with $\sqrt{S}=7~
{\rm TeV}$, the cross section of $t\bar t$ production at NLO QCD level in SM is $\sigma_{t\bar
t}=154.5 ~{\rm pb}$. Including NP contributions at the NLO BFP, we have $\sigma_{t\bar t,{\rm NLO}}^{\rm NP + SM}= 175 ~{\rm
pb}$.

From the shape of contours of $\sigma_{t\bar{t}} $ and $ \rm A_{\rm
_{C}}$, one may easily distinguish NP events from SM ones. The
location of the brown contours indicate that vector current $\rm
C_{\rm V}$ or axial currrent $\rm C_{\rm A}$ alone cannot improve
the quality of fit significantly. The acceptable confidence region
(global 95\% CL) centers around four points in the parameter space
where $|\rm C_{\rm V}|$ equals $|\rm C_{\rm A}|$, which means
experimental data favor purely right-handed or left-handed couplings.

\section{Conclusion}\label{s5}
 We have investigated NLO QCD effect on total cross section, invariant mass distribution and forward-backward asymmetry $\rm A_{\rm {FB}}$ of top quark
pair production  mediated by $W'$ at the Tevatron and LHC.
We have taken into account the interference of NP channel  with QCD channel ({\rm~up~to~} $\mathcal{O} ( \alpha_{s}^2 g'^{2})$),
as well as the interference  between NP channels({\rm ~up~to~}$\mathcal{O}( \alpha_{s} g'^{4})$).
We fit the data at the Tevatron, including total cross section, the invariant mass distribution and the $\rm A_{FB}$, and find the allowed parameter space. We show that due to the cancelation between these two parts of contributions,
the NLO total cross section exhibits only a slight modification compared to the LO result of NP.
But the  $\rm A_{\rm {FB}}$ is increased by about 9\%.
Moreover, NP couplings is constrained strongly by the discrepancy between NP and SM in the invariant mass distribution. And these constraints become more stringent in the QCD NLO.
Thus, it is difficult to satisfy simultaneously the constraints
from data of $ \rm A_{\rm FB}$ and $d\sigma / d m_{t \bar t}$ spectrum in the large invariant mass region at the Tevatron.
At the LHC, both total cross section and $A_{\rm C}$ can be used to distinguish NP from SM and therefore
the LHC may detect these NP effects with the integrated luminosity increasing.

\begin{acknowledgments}
This work was supported by the Undergraduate Research Fund of Education Foundation of Peking University
and National Natural Science Foundation of China, under Grants No.~11021092 and No.~10975004.
\end{acknowledgments}

\bibliography{Wprime}

\end{document}